*Article*

# Influence of Heat Accumulation on Morphology Debris Deposition and Wetting of LIPSS on Steel upon High Repetition Rate Femtosecond Pulses Irradiation

**Camilo Florian [1],\*, Yasser Fuentes-Edfuf [1], Evangelos Skoulas [2], Emmanuel Stratakis [3], Santiago Sanchez-Cortes [3], Javier Solis [1] and Jan Siegel [1]**

[1] Instituto de Óptica (IO-CSIC), Consejo Superior de Investigaciones Científicas (CSIC), Serrano 121, 28006 Madrid, Spain
[2] Instituto de Estructura de la Materia (CSIC), Consejo Superior de Investigaciones Científicas, Serrano 121, 28006 Madrid, Spain
[3] Institute of Electronic Structure and Laser (IESL), Foundation for Research and Technology (FORTH), N. Plastira 100, Vassilika Vouton, 70013 Heraklion, Greece
\* Correspondence: camilo.florian@csic.es

**Abstract:** The fabrication of laser-induced periodic surface structures (LIPSS) over extended areas at high processing speeds requires the use of high repetition rate femtosecond lasers. It is known that industrially relevant materials such as steel experience heat accumulation when irradiated at repetition rates above some hundreds of kHz, and significant debris redeposition can take place. However, there are few studies on how the laser repetition rate influences both the debris deposition and the final LIPSS morphology. In this work, we present a study of fs laser-induced fabrication of low spatial frequency LIPSS (LSFL), with pulse repetition rates ranging from 10 kHz to 2 MHz on commercially available steel. The morphology of the laser-structured areas as well as the redeposited debris was characterized by scanning electron microscopy (SEM) and µ-Raman spectroscopy. To identify repetition rate ranges where heat accumulation is present during the irradiations, we developed a simple heat accumulation model that solves the heat equation in 1 dimension implementing a Forward differencing in Time and Central differencing in Space (FTCS) scheme. Contact angle measurements with water demonstrated the influence of heat accumulation and debris on the functional wetting behavior. The findings are directly relevant for the processing of metals using high repetition rate femtosecond lasers, enabling the identification of optimum conditions in terms of desired morphology, functionality, and throughput.

## 1. Introduction

Laser-induced periodic surface structures (LIPSS) can be used for applications in areas such as optics, biology, and engineering [1]. In general, their fabrication process constitutes a single-step approach that can be performed under ambient conditions, ideal for most widespread industrial applications that incorporate lasers for micromachining purposes [2]. LIPSS formation occurs upon interference of the incident laser beam with the electromagnetic surface wave that is generated by it at the surface of the irradiated material, leading to a local intensity modulation pattern [3]. When the structures period is close to the laser wavelength, they are classified as low spatial frequency LIPSS (LSFL), and when the period is at least a factor two smaller, as high spatial frequency LIPSS (HSFL) [4]. For the so called LSFL, this periodic intensity translates into an ablation pattern of parallel structures with spatial dimensions ranging from nanometers up to several microns. Although the process is material-dependent [5], the study of the main processing parameters that lead to the formation of LIPSS have been identified for various metals, semiconductors and dielectrics [1], including material properties (optical [6,7], thermal [8], chemical [9]), processing conditions (number of effective pulses per spot unit [10–12], scanning direction vs. polarization [13], over-scanning [14], irradiation atmosphere [15], surface polishing [16], substrate temperature [17,18], material thickness [19,20]) and laser source parameters (wavelength [21], pulse duration [22], beam polarization [23], angle of incidence [24], and number of beams [25,26]). Lately, the increasing availability of affordable and stable high-repetition rate femtosecond lasers have been placed under the spotlight, with the laser repetition rate as a key processing parameter to investigate in detail [27–33]. One reason is because materials irradiated at such high rates present heat accumulation effects that can affect dramatically the overall formation and performance of LIPSS [32,34–38]. At the same time, the achieved high processing throughputs are accompanied by higher ablation rates, generating considerable amounts of debris redeposition. The latter may lead to significant performance changes of the fabricated structures, depending on the characteristics of the redeposited particles. Still, there are only relatively few studies that relate them with functional changes on the produced structures [27,39].

In this paper, we present a study showing the formation of low spatial frequency LIPSS (LSFL) in two types of commercially available steel with similar elemental compositions (see Materials and Methods section). We dedicated special attention to the morphological changes induced in LIPSS at different repetition rates, i.e., enabling different heat accumulation regimes, and their influence on the redeposited debris morphology. Scanning electron micrographs were recorded to characterize the morphology of the different irradiated areas, including estimations of LIPSS periodicities. Wettability measurements with water were performed on the irradiated areas in order to determine the influence of the repetition rate on the wetting behavior of the fabricated structures. Finally, we developed a simple thermal model that solves the heat flow equation in 1D using forward differencing in time and central differencing in space (FTCS) simulations. From a direct comparison of the simulation results with a set of experiments where low-spatial frequency LIPSS are formed, a defined range of operational repetition rates was determined, at which the influence of the heat accumulation on the LIPSS morphology in steel substrates is negligible.

## 2. Result and Discussion

### 2.1. Role of Repetition Rate on Redeposited Debris Morphology

It is known that steel substrates irradiated by femtosecond lasers are often covered by debris redeposition in the form of nanoparticle agglomerates [27–30]. However, less attention is paid to the fact that this redeposition layer is often non-robust and can be detached from the surface, directly impacting the surface wetting performance, as reported recently in [27]. In order to further investigate the processing parameters and their relation to the deposited debris, irradiation experiments in form of single lines and areas have been performed on steel. Figure 1A displays SEM micrographs acquired at the center of the LSFL lines, accompanied by micrographs recorded at unexposed regions close to the LSFL lines borders, where redeposited debris is present. The laser fluence for this experiment was $\phi$ = 1.5 J/cm$^2$ whereas the number of effective pulses, i.e., number of pulses per unit area, was kept constant at $N_{eff\ 1D}$ = 40, providing suitable conditions for LIPSS formation. A sketch of this experiment is included in Figure 1B, illustrating the incoming laser beam that induces the formation of LSFL along a line, and generating at the same time debris that is redeposited beyond the regions containing LIPSS, covering unexposed areas.

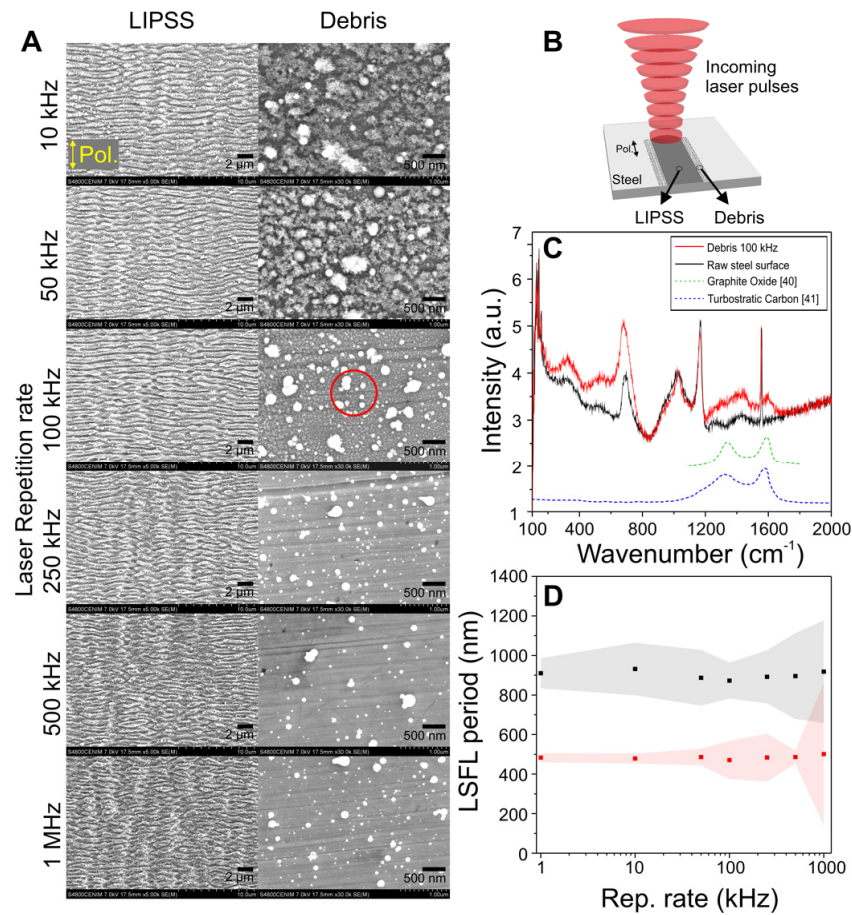

**Figure 1.** (**A**) SEM micrographs of LIPSS structures and redeposited debris in steel 1.7131, fabricated using a high laser fluence $\phi$ = 1.5 J/cm², and $N_{eff\,1D}$ = 40. Beam polarization is indicated by the double yellow arrow; (**B**) Sketch that indicates the scanning direction and the approximate locations where the images for the LIPSS and debris where taken; (**C**) Micro Raman characterization of the debris-region fabricated at 100 kHz (red circle in A) compared to the clean unirradiated steel surface and to data for graphite oxide and turbostratic carbon from references [40] and [41], respectively. Re-printed (adapted) with permission from Ref. [40]. Copyright 2008 American Chemical Society. (**D**) Plot of the LIPSS periods obtained from FFT maps (not shown) of the SEM micrographs. The major period is plotted in black squares, and the minor period in red circles. The shaded areas correspond to error bars calculated as indicated in Materials and Methods section.

At both sides of the structured lines, next to the LIPSS, SEM micrographs acquired at higher magnification show the redeposited debris in Figure 1A. Its morphology at lower repetition rates corresponds to dust-like material that covers almost completely the unexposed steel surface. It is important to differentiate this type of debris from the one redeposited on top of the LSFL, since particles redeposited there, after one pulse, will be subsequently irradiated by $N_{eff\,1D}$ number of pulses, consequently changing the debris morphology. As the repetition rate increases, the morphology slowly evolves towards a combination of well-defined spherical particles surrounded by dust-like material, the latter decreasing in density and exposing the polished unmodified steel surface underneath. At the highest repetition rates studied, dust-like particles are not present anymore and the debris consists of a low density of small spherical particles, leaving most of the underlying steel surface uncovered. This change of the debris morphology, especially at higher repetition rates (>250 kHz), i.e., where subsequent pulses arrive at the sample in short times (<5 μs), can be understood when considering the presence of clouds of ejected material transiently shielding the irradiated area, with lifetimes of several microseconds [34]. As a consequence, the probability for incoming pulses to interact with dust-like particles suspended in air increases with repetition rate. Possible effects of this interaction are that they are further defragmented, pushed back towards the surface and/or re-melted to form spherical particles upon solidification at the surface.

To confirm that debris particles have the same composition as the bulk sample, we have performed Micro-Raman spectroscopy on a non-irradiated region of the steel surface and in the region of the debris for the line fabricated at 100 kHz, as indicated by the red circle in Figure 1A. It can be concluded that the main features are essentially the same when comparing the Raman spectra plotted in Figure 1C. Of particular interest are the peaks located between 1300 cm⁻¹ and 1600 cm⁻¹. For the clean sample, the two peaks observed correspond to the vibrational modes D (1355 cm⁻¹) and G (1581 cm⁻¹) of crystalline carbon, present in the steel sample. Interestingly, two additional broad peaks can be observed in the debris region, centered at around 1400 and 1600 cm⁻¹. There are two possible origins for these peaks. First, and most likely, they are due to the formation of graphite oxide [40] since a reaction of the ejected material with the oxygen of the processing atmosphere is very likely. A second possibility, consistent with these peaks, is the formation of turbostratic carbon [41,42] that has a structural ordering in between amorphous carbon and crystalline graphite. The first contains varying quantities of $sp^3$ hybridized carbon atoms and the second, $sp^2$ hybridization [41,42]. If present, the formation of this structure likely occurs also during the material ablation.

## 2.2. Energy Absorption on Material Cloud Particles

The influence of the repetition rate via cloud shielding on the energy absorption at the steel surface was studied in the following experiment. Single near-threshold ablation lines were fabricated at repetition rates ranging from 1 kHz to 1 MHz, with a constant number of effective pulses per spot unit of $N_{eff\ 1D}$ = 10 and peak fluence $\phi$ = 225 mJ/cm² (above the fluence threshold for multi-pulse ablation $\phi_{th}$ = 164 mJ/cm²). The fluence value was chosen near threshold in order to increase the sensitivity to small energy variations. Figure 2A displays optical microscopy images of the different lines written, showing a significant decrease in line width and contrast for high repetition rates. Importantly, the contrast of these images was normalized via software to an unirradiated area using the same illumination and magnification conditions. Assuming the creation of a particle cloud suspended close to the sample surface with lifetimes of about ~5 μs, the particle cloud absorbs a significant fraction of the energy from the subsequent pulses, particularly at 1 and 2 MHz where the inter-pulse time is between 0.5 to 1 μs. A more quantitative analysis of the results is displayed in Figure 2B, showing reflectivity cross sections of these images. Since the lines borders are not perfectly defined, we implemented a threshold based on a relative normalized reflectivity in the range of 0.9 to 0.91. Within this range, we measure the line widths, and we plotted them in Figure 2C as a function of the repetition rate used in each case. From these results, at repetition rates lower than 500 kHz, both the width and contrast are similar. Starting at 500 kHz, the line width decreases dramatically, which is consistent with the scenario of a material cloud partially shielding the steel surface via absorption/scattering of a small fraction of the incident energy. This process most likely induces particle fragmentation and/or melting, which is responsible for the effects shown on the debris displayed in Figure 1.

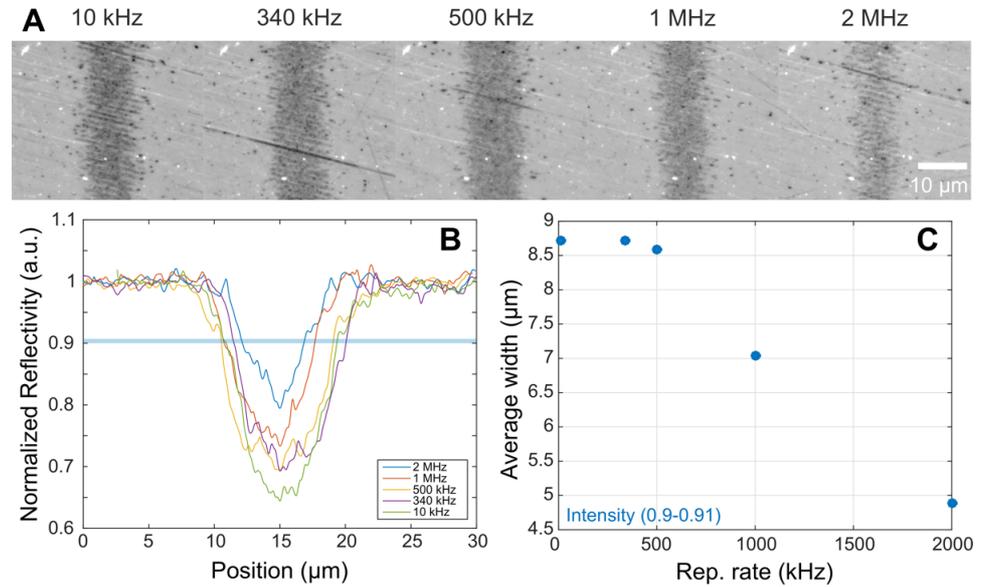

**Figure 2.** (**A**) Optical microscopy images of regular ablation lines ($N_{eff\,1D}$ = 10) produced at different repetition rates and constant near-threshold fluence $\phi$ = 225 mJ/cm²; (**B**) Normalized reflectivity cross-sections of the lines shown in (**A**); and (**C**) Plot of the line widths measured within the reflectivity range 0.9 to 0.91 highlighted as a blue shaded line in (**B**).

### 2.3. Heat Accumulation Calculations from the FTCS Simulations

In general, when a bulk material is irradiated with an ultrashort laser pulse, part of the energy is reflected and part of it is absorbed by the material. The energy from the incoming photons is absorbed linearly if it is larger than the material's bandgap, or non-linearly if an intensity threshold is overcome due to the high photon density. The mechanisms of energy absorption with femtosecond pulses have been largely studied and are summarized in [43]. Once the energy is deposited in the electron system, it relaxes to the lattice via electron-phonon coupling and conventional heat diffusion takes place. The speed at which this process takes place varies among materials, being generally fast for metals due to their high thermal conductivity. When the irradiation is performed with a train of pulses that are sufficiently close in time, i.e., at high repetition rates, the material might not have enough time to cool down completely, leading to an increase of the base material temperature. This is what is known as heat accumulation. Different models have been developed in the past, where the heat equation was solved numerically for heat accumulation in silicon [31], steel [38] and borosilicate glass [44]. One common feature of these models is that they do not consider phase transitions happening at temperatures above the melting point. However, as a first approximation, they allow for the identification of suitable ranges of repetition rates where the material exhibits strong heat accumulation.

In order to correlate these morphological changes with possible heat accumulation effects at certain repetition rates, we have implemented a simple model that implements a forward differencing in time and central differencing in space (FTCS) scheme to solve the heat flow equation in 1-dimension. It considers a 1-dimensional solid steel bar of length $l$ = 1 mm at room temperature, heated by a Gaussian temperature distribution with the same geometry and fluence as used during the experiments, employing the temperature dependent thermal properties of steel 1.7131. The FTCS simulates an experiment with 500 pulses (unless indicated otherwise) delivered to the sample under the same conditions of the experiment (Gaussian beam geometry, repetition rates and effective number of pulses per spot). Details on the specific operating modes (*static probe* and *dynamic probe*) are included in the Materials and Methods section. The results of the simulations are presented in Figures 3 and 4.

Initially, the model was used in *static probe* mode to quantify the linear heat dissipation velocity in the material. To this end, the time-dependent temperature rise and decay was probed at a fixed position ($x = 0$ μm, the center of the 1D bar) and single pump pulses heat the material at different distances from the probe ($x = 30$, 100 and 200 μm), as shown in Figure 3A. As expected, the closer the probe to the pump, the higher the amplitude of the temperature rise and the earlier it appears. From the obtained plots, the time difference between the pulse arrival and the peak temperature at the temperature probe position can be calculated directly, which enables a quick calculation for the heat dissipation speed along the bar. Additional positions are plotted in Figure 3B, where time is plotted versus position in order to directly compare with the corresponding dissipation speeds at the same positions. When the pulse lands directly on the probe position (for distances smaller than the laser beam waist, in our case ~20 μm), the calculated times are not included since they would not be meaningful for the study of lateral heat flow (shaded area in Figure 3B).

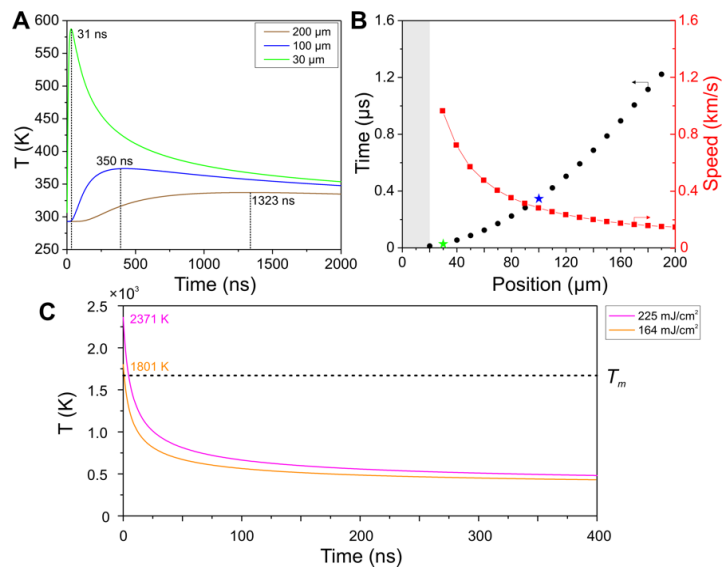

**Figure 3.** (**A**) Temperature curves in *static probe* mode evolving over time for a single laser pulse $\phi_{th}$ = 164 mJ/cm² arriving at different positions $x$ (30, 100 and 200 μm) and sensed by the temperature probe fixed at $x = 0$ μm. The inset numbers indicate the time that took the laser-deposited heat to reach the position of the probe and register the peak temperature; (**B**) Time that takes the heat to travel to a position $x = 0$ μm vs the position where the laser pulse impinged on the bar. The times for positions 30, 100 and 200 μm from (**A**) are plotted as green, blue and gold star symbols, respectively. The additional axis at the right represents the speeds (x/t, in km/s) at which the temperature maxima are being detected. The shaded area corresponds to the position of the beam waist, for which times were not measured. Data plotted on black circles use the Time (μs) axis, whereas data on red squares and red dashed line use the Speed (km/s) axis, as indicated by the small inset arrows; (**C**). Simulation of a single pulse impinging on the position of the temperature probe ($x = 0$ μm) with fluences $\phi = 225$ mJ/cm² and 164 mJ/cm² without considering phase transformation effects. The maximum temperatures registered are 2371 K and 1801 K, respectively. The dash line at 1672 K corresponds to the melting temperature for steel.

A second simulation in *static probe* mode, with a single heat pulse at the position of the temperature probe ($x = 0$ μm) at two different fluences ($\phi = 225$ and 164 mJ/cm²), is included in Figure 3C. The maximum temperatures registered are 2371 K and 1801 K, respectively. In particular, the simulation shows that that at the higher fluence used, which was identified experimentally as the threshold for single-pulse ablation, the calculated maximum temperature rise upon single pulse irradiation significantly exceeds the melting temperature for steel 1.7131 ($T_{melt}$ = 1672 K). In the case of the lower fluence identified as the threshold for multi-pulse ablation, the temperatures are very similar (again, without considering any material phase changes). Taking into account that, in general, the ablation threshold for multi-pulse irradiation (even when the pulse repetition rate is well below the thermal accumulation regime) is considerably lower than the single-pulse ablation threshold, this result can be considered as a solid confirmation of the model in terms of absolute temperature increase calculated for a given fluence.

Figure 4 displays the simulation results operating in *dynamic probe* mode, where a local temperature probe follows the landing position of laser pulse, with a fast-moving pump laser beam at $\phi$ = 0.5 J/cm², starting at $x$ = −200 μm and advancing quickly through the bar. Each curve in Figure 4A displays laser-induced temperature pulses with peaks that correspond to the arrival time of pulses to the sample, sensed at the pulse landing position. For example, at 50 kHz, each pulse reaches the sample surface with an interpulse time of 20 μs. For each repetition rate, the laser beam scanning speed is adjusted to keep the same effective number of pulses as $N_{eff\,2D}$ = 30. For details on the simulation configurations used, see Materials and Methods section. The strong effect of heat accumulation can be seen by the fact that the peak amplitude clearly increases with repetition rate. The underlying reason is that for high repetition rates the material does not have sufficient time to cool down before a subsequent pulse reaches the sample. This effect can be clearly appreciated not only by the increase of the peak temperature but by an increase of the base temperature.

Figure 4B,C display the evolution of the peak and base temperatures respectively achieved for different fluences as a function of the repetition rate. For all fluences, both, the peak and base temperatures remain almost constant upon a repetition rate increase from 10 kHz to 100 kHz. However, from 250 kHz onwards, a significant increase of both temperatures can be observed, and the amount of increase depends strongly on the laser fluence. In particular, for 0.5 J/cm², the base temperature reaches a value of ~3200 K at 2 MHz. This value lies significantly above the melting temperature, which implies that the subsequent pulse is not only incident onto a hot material but onto the molten phase. An even more extreme scenario can be found for 0.75 J/cm², where the base temperature reaches a value of ~11200 K at 2 MHz, significantly above the evaporation temperature.

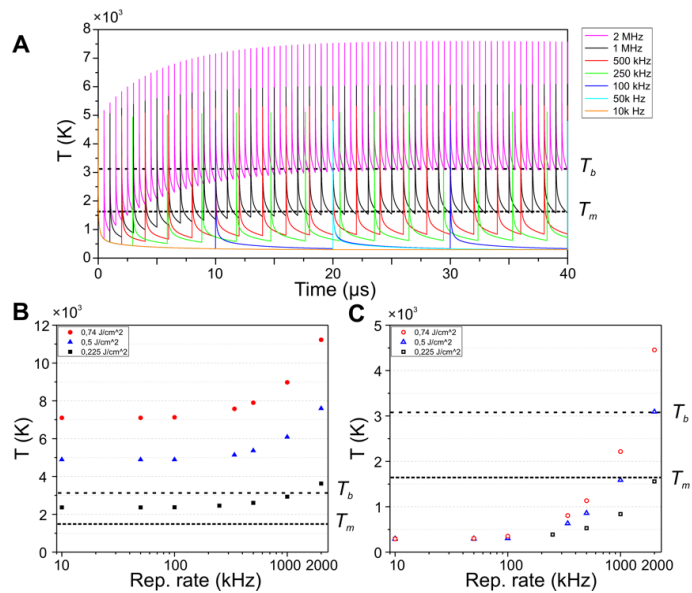

**Figure 4.** (**A**) Temperature curves versus time for different repetition rates acquired in *dynamic probe* mode, showing multiple peaks that correspond to the arrival of a pulse over the moving

temperature probe. The experimental parameters are the same of the experiment shown in Figure 5A ($\phi$ = 0.5 J/cm², and $N_{eff\,1D}$ = 30); (**B**) peak temperature; and (**C**) base temperature from simulations for different repetition rates. The dashed lines in correspond to melting and boiling temperatures of steel of 1672 K [45] and ~3300 K [46].

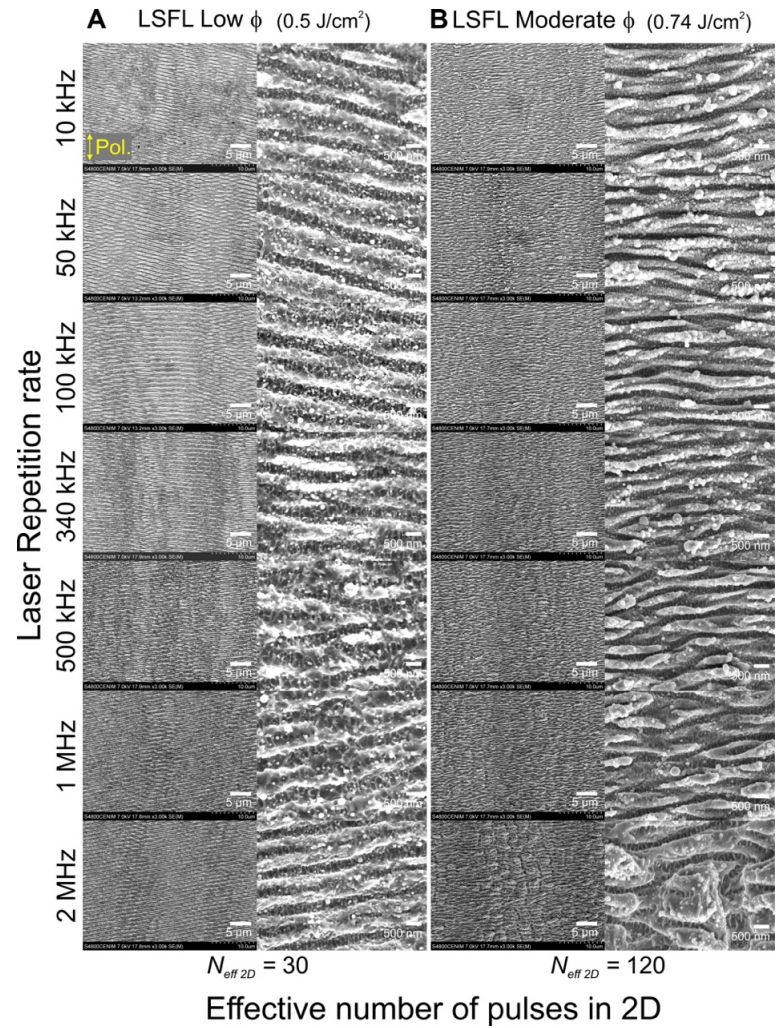

**Figure 5.** Areas irradiated on steel 1.7225 at different laser repetition rates. The structures shown are organized in two vertical sets (and two different magnifications) corresponding to LSFL at low (**A**) and moderate (**B**) fluences. Each set of structures is produced at a fixed laser fluence ($\phi$) and a constant number of effective pulses per spot unit in 2-dimensions ($N_{eff\,2D}$) indicated at the bottom of each structure set. The laser beam polarization is indicated by the yellow double arrow in the top left SEM micrograph. The laser scanning direction is parallel to this direction.

At this point, according to the information in Figure 4B and Figure 4C, we can conclude that heat accumulation is present for repetition rates in the range of 250 kHz to 1 MHz and it is more pronounced with a fluence increase. As it will be shown in the next section, this result explains the morphology changes of the LSFL.

### 2.4. Low Spatial Frequency LIPSS Areas

A general morphological characterization of the irradiated regions in steel 1.7225 for different pulse repetition rates is included in Figure 5 via SEM micrographs. The first set of experiments consisted in the fabrication of two different sets of structures (vertical columns in Figure 5) including LSFL obtained at low and moderate fluences. Importantly, the effective number of pulses per line ($N_{eff\,2D}$) was kept constant for the different repetition rates explored by adjusting accordingly the scanning speed, as indicated in the Materials and Methods section. Within the laser processed areas, SEM images at two magnifications were recorded, one at low magnification that shows the parallel periodic structures (LSFL), and a second one at higher magnification that reveals the presence of redeposited nano-scale debris on the LSFL.

For the case of LSFL at low fluences in Figure 5A, well-pronounced ripple structures are visible. The general structure orientation follows a trend perpendicular to the polarization, however, due to the low fluence and number of pulses used ($\phi = 0.5$ J/cm², $N_{eff\,2D} = 30$), this expected orientation is affected by the natural orientation of the polished surface scratches, resulting in orientations that are slightly off the horizontal expected direction. This effect is more visible in the optical microscopy images of Figure 2, where the natural scratches on the unirradiated sample dominate and define the final orientation of the LSFL. For the areas in Figure 5B irradiated at moderate fluence ($\phi = 0.74$ J/cm²) and a higher number of effective pulses ($N_{eff\,2D} = 120$), the influence of the natural scratches is overcome and the LSFL are oriented horizontally with no offset inclination. At this fluence, the LSFL produced in the range of 340 kHz to 2 MHz, present visible deformations corresponding to ripples with slightly larger periods and less debris superimposed. These results agree well with our model in Figure 4 regarding the repetition rate range in which the morphology start to show visible changes in Figure 5B. Additionally, the formation of grooves parallel to the polarization [4,47] with periods around ~3.0 μm that break the continuity of the horizontal ripples are present, more visible for the structures produced at 2 MHz. The presence of grooves, not detected for the LSFL at low fluences, suggest that at this fluence and effective pulse number, the morphology of the LIPSS start to present features produced by thermal effects at high repetition rates.

A second higher magnification of the irradiated areas is included in Figure 5. These SEM micrographs allow a detailed inspection of the structures and the redeposited debris on top of them. For the LSFL at low fluence, areas irradiated at low repetition rates show the redeposition of dispersed nanoparticles with spherical shape and diameters around ~150 nm (tiny white dots on top of the LSFL structures). Their size slightly increases with the repetition rate up to ~300 nm at 340 kHz, whereas they almost disappear at 2 MHz. For the LSFL at moderate fluence, the nanoparticles form agglomerates and are homogeneously distributed over the LSFL, featuring fewer clusters of spherical particles with diameters of ~400 nm for the lower repetition rates, and decreasing in number and size for higher repetition rates.

## 2.5. Period Splitting of LIPSS

In order to inspect the LIPSS morphology in more detail, a powerful tool to quantitatively measure both the structures orientation and period consists of performing a fast Fourier transform (FFT) on the SEM micrographs of Figure 1 and Figure 5. The results obtained from this operation are so-called FFT maps that contain a cloud distribution of points in the Fourier space, which—for periodic structures—feature high density regions at positions in the reciprocal space that indicate their period and orientation in the spatial domain (for details, see Materials and Methods section). Figure 1C displays the different periods for each repetition rate. From there, two average periods for structures perpendicular to the laser polarization of $900 \pm 150$ nm and $485 \pm 100$ nm are present in the irradiated regions. Similar to the results in Figure 1, the periods of the LSFL at low fluences ($\phi = 0.5$ J/cm², $N_{eff\,2D} = 30$, Figure 6A) present two different average periods that remain practically the same, regardless the repetition rate used: a major one at $1010 \pm 110$ nm close to the laser wavelength ($\lambda = 1030$ nm), and a minor one at $510 \pm 55$ nm (black and red squares, respectively). For the LSFL at moderate fluences ($\phi = 0.74$ J/cm², $N_{eff\,2D} = 120$, Figure 6B), the FFT maps also reveal the presence of two periods at $950 \pm 150$ nm and $490 \pm 110$ nm, except for the one obtained at 2 MHz where only a major period can be detected.

The occurrence of this period splitting has been reported before in cooper and steel samples [47–49]. In these works, the splitting was explained by a mechanism taking place when the ripples have achieved a certain depth, at which the coupling between the surface plasmons and the incident laser radiation enables a feedback process in which a local enhancement developed at the center of the LSFL, effectively triggering ablation and thus splitting of the ripples. In our experiments, this mechanism translates into period splitting around $\lambda/2$, confirmed by the FFT maps included in Figure 6C,D. These results demonstrate that period splitting can be a rather stable and reproducible process, feasible at different fluences and repetition rates. Overall, the repetition rate and redeposited debris does not appear to significantly modify the LIPSS period. For repetition rates equal or higher than 250 kHz, supra-wavelength structures with a period of ~2.7 µm appear, aligned parallel to the laser polarization, which correspond to grooves [4,47] also observed in area scans.

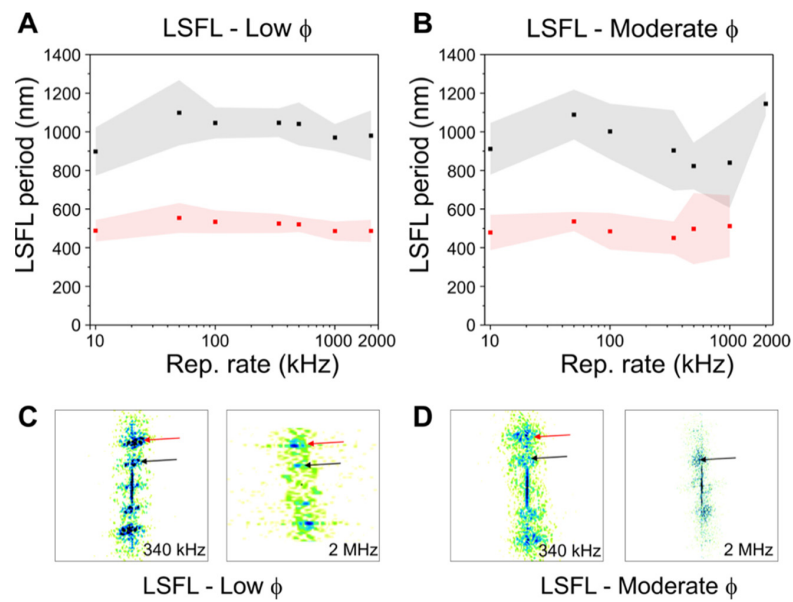

**Figure 6.** LIPSS periodicity at (**A**) low and (**B**) moderate laser fluences (ϕ) for different repetition rates, ranging from 10 kHz to 2 MHz. The LSFL structures contain two sets of periodicities, one plotted as black squares (and gray error bars as shadowed areas) for periodicities close to the laser wavelength (~1 µm) and a second one plotted as red squares (and red error bar regimes as shadowed areas) for smaller periods (~500 nm). The central period values and error bars were calculated from the FFT maps following the procedure described in [24] for each SEM micrograph shown in Figure 5. A selection of FFT maps is included in (**C**) for LSFL at low fluences at 340 kHz and 2 MHz, and (**D**) for moderate fluences for the same repetition rates, to visualize the coexistence of two clouds of points, indicative of the presence of two different periods, except for LSFL—moderate fluence at 2 MHz.

### 2.6. Wettability Behavior of LIPSS

In the past, it has been demonstrated that surface wetting on laser-irradiated samples depend not only on the surface morphology, but also on processing parameters that affect the chemistry at the surface [50,51]. For this reason, in order to estimate the influence of the repetition rate on the functionality of the produced structures, we characterized the irradiated areas by performing contact angle (CA) measurements of a water drop placed onto the laser-processed surface, following the procedure described in the Materials and Methods section. All areas presented in Figure 5 were characterized 15 days after irradiation, in order to allow the sample to chemically stabilize to obtain constant CA values [27,37]. The CA values are plotted vs the repetition rate in Figure 7 for all structures fabricated. For the LSFL obtained at low fluence (Figure 7A), the surfaces are hydrophobic (CA > 90°), ranging between 110°–130°, with the CA slightly decreasing with repetition rate. The slightly higher CA values at lower repetition rates can most likely be attributed to additional roughness provided by the more abundant redeposited debris that facilitates the creation of packed air pockets influencing the overall surface wetting behavior [11,27]. For the LSFL obtained at high fluence (Figure 7B), the surface wetting behavior is also hydrophobic. The maximum CA value ($CA_{max} = 150°$) is reached for the areas produced at 340 kHz, after which the wettability slightly decreases. This result might be related to both, the appearance of the morphology deformations for the structures fabricated at repetition rates of 340 kHz and above as shown in Figure 5.

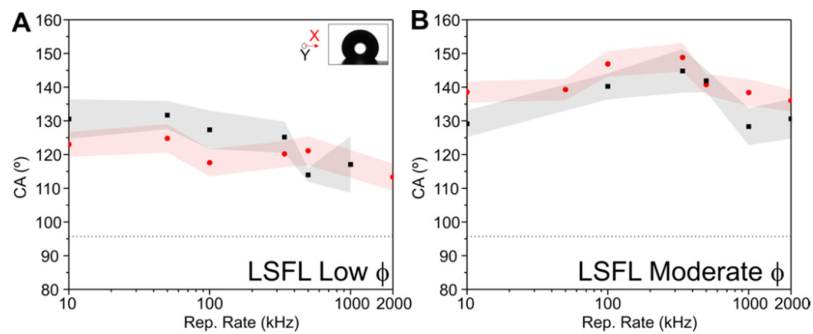

**Figure 7.** Contact angles (CA) measured from two different viewing positions X (red dots) and Y (black squares) of laser irradiated areas (droplet image included in the inset) for LSFL at (**A**) low and (**B**) moderate fluences. The horizontal dotted line corresponds to the CA measured on the un-irradiated sample surface for comparison. The error bars displayed as shadowed areas were determined as indicated in the Materials and Methods section.

## 3. Conclusions

The fabrication of LIPSS with high repetition rate femtosecond lasers produces significant heat accumulation effects that affect not only the morphology but also the functionality of the fabricated structures. In the particular case of steel, our experimental results supported by a heat accumulation model predict a range of repetition rates (>200 kHz) where the effects of repetition rate are negligible on the morphology of the LIPSS and the redeposited debris, however, significant morphology changes can be associated with irradiations at repetition rates above 200 kHz, which are in line with the simulated temperature increase due to heat accumulation. Chemical characterization via Raman spectroscopy reveals that the redeposited debris composition sustains certain degree of oxidation in form of graphene oxide and/or turbostratic carbon. Finally, the wetting characterization of LIPSS produced under different repetition rates shows slight wettability changes associated to morphological changes on the LIPSS induced by heat accumulation occurring at repetition rates above 200 kHz.

## 4. Materials and Methods

**Materials.** The materials used for the irradiation experiments were commercially available steel 1.7131 (16MnCr5—Fe:97.4%, C:0.16%, Si:0.25%, Mn 1.15%, Cr:0.95%, S:0.035%) and steel 1.7225 (42CrMo4—Fe:97.5%, C:0.38%, Si:0.4%, Mn:0.6%, Cr:0.9%, S:0.035%, Mo:0.15%, P:0.025%). The elemental composition of both steel types present a concentration of ~97.5% iron, thus providing overall similar material properties, including the thermal conductivity at room temperature (~43 W/(m.K)) and specific heat capacity (~460 J/(kg·K)), according to the available data sheets [52–55]. The surfaces in both cases were mirror-like polished, obtaining a roughness of $R_a$ < 10 nm. The cleaning of the samples before and after irradiation was done by an ultrasonic bath in acetone for 5 min, gently dried afterwards to eliminate the residues, except for the experiment on debris characterization shown in Figure 1, for which no cleaning was performed after laser irradiation to keep unattached nanoparticle agglomerates from falling off the surface. The samples were stored in a desiccator, keeping a relative humidity of 30%.

Laser setup. The implemented laser setup consisted in a high repetition rate femtosecond laser (model Satsuma, Amplitude Lasers, Pessac, France), delivering pulses of 350 fs with central wavelength of 1030 nm. The active media is an Yb-doped fiber, offering nominal output power of 20 W at 500 kHz. The operating repetition rates used during the experiments ranged from 10 kHz up to 2 MHz. The energy of the pulses was controlled by a combination of a half-wave plate and a thin film polarizer. The beam was scanned over the sample surface in the XY plane by means of a scanning head from LaserScan Scanlab III® (SCANLAB GmbH, Puchheim, Germany) and the pulses where focused to a beam waist $w_0$ = 21.15 μm (calculated via Liu method [56]) with a F-theta lens of 100 mm focal length and a usable field of view of $7 \times 7$ cm². A sketch of the system can be found in [51].

Irradiation strategies. Irradiations were done to fabricate lines and areas. For the 'lines' experiments, or 1-dimensional modifications, the procedure consisted in spatially overlapping pulses to form continuous lines. The effective number of pulses is calculated considering the geometry of the focused beam and the processing parameters as $N_{eff\,1D}$ = $2w_0 f / V$, where $2w_0$ is the beam diameter, $f$ the laser repetition rate, and $V$ the scanning speed. Such structures are displayed in Figure 1, $N_{eff\,1D}$ = 10 and Figure 2, $N_{eff\,1D}$ = 40. For the '*areas*' experiments, or 2-dimensional modifications, $0.5 \times 0.5$ cm² squares were irradiated by overlapping consecutive lines to produce a rather homogeneous structured area. The effective number of pulses is given by $N_{eff\,2D} = N_{eff\,1D} \times \Delta = 2w_0 f / V \times \Delta$, where $\Delta$ is the interline separation between two consecutive lines [21,43]. The parameters for the structures in Figure 5 are at low fluence, $N_{eff\,2D}$ = 30 and $\Delta$ = 20 μm, and for moderate fluence, $N_{eff\,2D}$ = 120 and $\Delta$ = 26 μm.

Surface characterization. After irradiation, the morphological characterization was done by optical microscopy for Figure 2 and scanning electron microscopy (SEM, Hitachi S-4800, Tokyo, Japan) for Figure 1 and Figure 5. A Fast Fourier Transformation of the LSFL structured areas was performed to determine the periodicity of the structures on the SEM micrographs. Central periodicity values and corresponding errors associated were obtained following the procedure described in [24]. Selected samples were characterized via μ-Raman spectroscopy (Renishaw InVia, Renishaw plc, Wotton-under-Edge, UK) to identify the possible compositional changes of the sample and the produced debris under different irradiation conditions. The spectra shown are the result of an average of 5 acquisitions using a laser power at the sample of 1 mW at 532 nm wavelength. The light was focused with a 100× objective with numerical aperture NA = 0.85 down to a spot of ~1 μm in diameter.

Wetting characterization. Contact angle (CA) measurements were performed on an OCA 15EC system (DataPhysics Instruments GmbH, Filderstadt, Germany) equipped with a CCD camera to capture lateral images of a 3 μL deionized water droplet deposited on the selected laser irradiated region of the steel samples. The camera, the droplet, and the back-side illumination source were aligned such that the deposited droplet projects a contour image on the imaging camera. CA values as well as measured errors were obtained via software analysis after each measurement. All CA measurements presented in this work were performed at least 15 days after laser exposure [37].

Heat accumulation model. A simple model based on the "Forward differencing in Time and Central differencing in Space" FDTC scheme is used to find a numerical solution to the partial differential equation (1) for the heat flow in 1 dimension. This scheme has been applied in the past for heat diffusion problems when the energy source is computed as periodic Gaussian energy distributions attributed to laser pulses, as indicated in [31,38,44,57]. Importantly, the system evaluated here is not in thermal equilibrium since regardless of the used mode, there are always temperature variations as time increases. The simulation considers a 1-dimensional solid steel bar at room temperature, heated by a Gaussian temperature distribution with the same geometry and fluence as used during the experiments, employing the temperature dependent thermal properties of steel 1.7131 for the calculation of the heat dissipation coefficient. For the model validation and the calculation of the heat diffusion speeds in the material, the *static probe mode* was used consisting of a fixed temperature probe was placed at the center of the 1D bar positioned at $x$ = 0 μm, while single pulses heated the material at different positions, as shown in Figure 8. The simulation results from Figure 3A,B were obtained using a static temperature probe and a single pulse located at certain positions (30, 100 and 200 μm), following the scheme of Figure 8A. For the simulation results of Figure 4, the same processing parameters (scanning speeds, fluence and repetition rates) were used, using the *dynamic probe mode* consisting of a moving temperature probe and a scanning beam that moves along the 1D bar starting from $x$ = −200 μm at the set scanning speed, as indicated in Figure 8B. The temperature curve fed to the simulation every time a pulse reaches the bar correspond to a Gaussian distribution, as illustrated in Figure 8C. This temperature is added to the bar cells once a pulse reaches the sample. The simulation runs a FTCS approach (Forward differencing in Time and Central differencing in Space at time level $n$) for solving the 1-D heat equation. The initial temperature of the bar is room temperature (293 K). For simplification, the boundary conditions are constant at room temperature and the bar is set long enough to avoid any accumulation when approaching both ends. The heat transfer coefficient is calculated in each iteration on the simulations since it depends on the materials temperature. It is calculated in our code by considering the temperature dependent specific heat and thermal conductivity, found in [52,53] for steel 1.7225 and [54,55] for steel 1.7131. When simulated temperatures exceed the available data from [52–55], the simulation uses the highest experimentally reported values for the calculation of the coefficient. The temperature $T_{xi}$ for a unitary cell in a time $t_n$ is the average of the two neighbors' and the same cell at time $t_{(n-1)}$ as indicated in Figure 8D. In our particular case, $dx$ = 15 nm and $dt$ = 1 ns, which gave more reliable results when comparing heat accumulation results on Silicon using the same materials data and processing parameters reported in [31]. Specific heat equations and details on the method implemented for the simulation results are included in the Supplementary Materials.

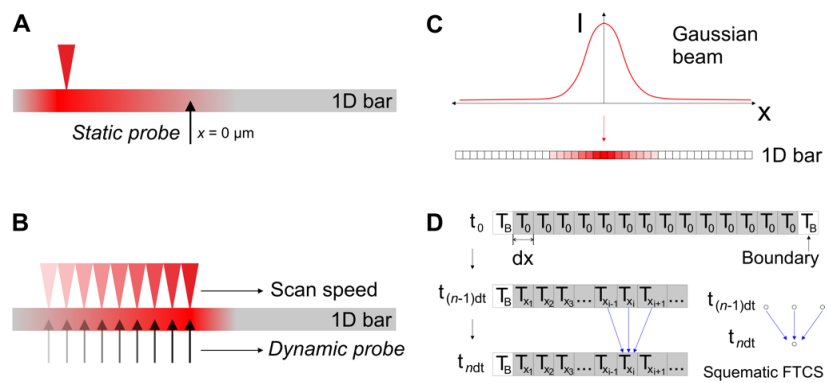

**Figure 8.** Schematic representation of the used model that implements the 1-D heat flow equation in a 1-D steel bar. Two general modes are employed, (**A**) a *static* probe that registers the temperature of individual pulses arriving at different positions from the probe used for simulation results of Figure 3A,B, and (**B**) a *dynamic* temperature probe and a scanning laser beam that heats the steel bar and moves from left to right at a given speed, such as in the experiments shown in Figure 3C. The moving temperature probe registers the temperature at intervals $dt$ = 1 ns. (**C**) The temperature curve distribution used as input for each pulse correspond to a Gaussian beam, translated into the 1D bar individual cells which size is $dx$ = 15 nm. (**D**) FTCS scheme for the temperature averaging of individual cell calculated every time increment $dt$ = 1 ns.

**Supplementary Materials:** The following supporting information can be downloaded at: https://www.mdpi.com/article/10.3390/ma15217468/s1: Detailed equations used for the forward differencing in time and central differencing in space (FDTC) scheme implemented for the simulations presented in Figures 3 and 4.

**Author Contributions:** C.F. wrote the original draft, performed the experiments, and characterized the samples. C.F. and J.S. (Jan Siegel) wrote the manuscript. C.F., Y.F.-E., J.S. (Javier Solis) and J.S. (Jan Siegel) analyzed the results, C.F. and Y.F.-E. performed the simulations, C.F., Y.F.-E. and J.S. (Jan Siegel) designed the experiments. E.S. (Evangelos Skoulas) and E.S. (Emmanuel Stratakis) performed the wetting characterization. S.S.-C. performed the Raman measurements. All authors contributed to the scientific discussion and revision of the article. C.F. and J.S. (Jan Siegel) directed the project. All authors have read and agreed to the published version of the manuscript.

**Funding:** This research was supported by the LiNaBioFluid Project (H2020-FETOPEN-2014–2015RIA, Grant No. 665337) of the European Commission as well as the research grants (TEC2017-82464-R and PiD2020-112770RB-C21) from the Spanish Ministry of Economy and Competitiveness. C.F. acknowledges the support from the European Commission through the Marie Curie Individual Fellowship—Global Grant No. 844977.

**Institutional Review Board Statement:** Not applicable.

**Informed Consent Statement:** Not applicable.

**Data Availability Statement:** The data presented in this paper is available upon request to the corresponding author.

**Acknowledgments:** Especial thanks to Antonio Tomás-López and Alfonso García-Delgado from CENIM-CSIC for the SEM characterization. Also, to Judith Swan and the team of the Writing in Science and Engineering program from Princeton University, for valuable and thoughtful comments on the structuring of this paper.

**Conflicts of Interest:** The authors declare no conflict of interest.